\documentclass{ws-p9-75x6-50}

\begin{document}

\title{Oscillatory approach to the singularity
in vacuum $T^2$ symmetric spacetimes}

\author{M. Weaver}

\address{Max Planck Institut f\"{u}r Gravitationsphysik,
D-14476 Golm,
Germany\\E-mail: weaver@aei-potsdam.mpg.de}

\author{B. K. Berger}

\address{Department of Physics,
Oakland University, Rochester, Michigan 48309,
USA\\E-mail: berger@Oakland.edu}

\author{J. Isenberg}

\address{Department of Mathematics,
University of Oregon, Eugene, Oregon 97403,
USA\\E-mail: jim@newton.uoregon.edu}

\maketitle

\abstracts{A combination of
qualitative analysis and numerical study indicates
that vacuum $T^2$ symmetric spacetimes are,
generically, oscillatory.}

Thirty years ago, Belinskii, Khalatnikov and Lifshitz
proposed that the dynamics of spatially inhomogeneous
solutions to Einstein's equation near a cosmological
singularity are oscillatory in general.\cite{bkl}
The quantities which oscillate are the
generalized Kasner exponents, which are the eigenvalues of the
extrinsic curvature divided by the mean curvature.  In various
known cases, there exists a foliation and threading of a
solution to Einstein's equation in the neighborhood of a
cosmological singularity such that the generalized Kasner
exponents along each thread either converge to a limit at
the singularity (if this occurs on all threads we
call the solution convergent),
or approximately follow the BKL sequence.\cite{bkl}
Along threads with oscillatory behavior, the BKL
sequence is realized because the evolution becomes approximately
Kasner, but then there is inevitably a transition to a
different Kasner evolution.  Each transition can be approximated,
yielding the BKL sequence.  The singularity is
at finite proper time, but there are an infinite number of
oscillations.

Numerical simulations of vacuum $T^2$ symmetric spacetimes
show that an ``asymptotic regime'' is
reached in which the generalized Kasner exponents follow
a portion of the BKL sequence.  Qualitative analysis
in the asymptotic regime indicates
that the oscillations will, in general, continue without end
at almost every spatial point.  These methods have been
described previously.\cite{pap1,pap2,U(1)}

Gowdy spacetimes are vacuum $T^2$ symmetric spacetimes
such that the (spacelike) symmetry orbits are
surface orthogonal.  It is thought that all Gowdy
spacetimes are convergent.  We find oscillatory dynamics in
non-polarized $T^2$ symmetric spacetimes if the symmetry
orbits are not surface orthogonal, in which case the
spatial topology must be the three torus.  A global foliation
and threading of these spacetimes is known.\cite{globalfol}
The metric can be written
\begin{equation}
\begin{array}{rcl}
g & = &  -e^{{\lambda - 3 \tau \over 2}} d \tau^2 +
e^{\lambda + \mu + \tau \over 2} d \theta^2
+ \sigma \, e^{P - \tau} [dx + Q \, dy +(G_1 + Q \, G_2)
\, d\theta\\[4pt]
&&{}- (M_1 + Q \, M_2) \, e^{-\tau} d\tau]^2
+ \sigma \, e^{-P-\tau} [dy + G_2 \, d\theta - M_2
\, e^{-\tau} d\tau]^2,\\[4pt]
\end{array}
\label{metric}
\end{equation}
with the singularity in the direction of increasing $\tau$.
The metric functions are independent of $x$ and $y$ and
periodic in $\theta$.  If $Q$ vanishes everywhere
the solution is polarized, and thought to be
convergent.\cite{pol}  If the non-positive function
$\partial_\tau \mu$ vanishes everywhere, the symmetry
orbits are surface orthogonal.
If $|\partial_\tau \mu| \ll 1$, then the generalized
Kasner exponents are approximately
\begin{equation}
\label{kasnerexp}
{\partial_\tau \lambda + \partial_\tau \mu + 1
\over \partial_\tau \lambda + \partial_\tau \mu - 3}
\hspace{40pt}
{2\,(v-1) \over \partial_\tau \lambda + \partial_\tau \mu - 3}
\hspace{40pt}{ -2\,(v+1 ) \over \partial_\tau \lambda
+ \partial_\tau \mu - 3}.
\end{equation}
The error is less than $30 \sqrt{- \partial_\tau \mu}$, which
we show studying perturbations of linear operators.
The denominator in~(\ref{kasnerexp}) is strictly negative.

Numerical simulations of a spacetime with metric~(\ref{metric})
show that the evolution becomes approximately Kasner at each
value of $\theta$.  The signature for this is that both $\mu$
and also
$v = \sqrt{(\partial_\tau P)^2 + e^{2 P}(\partial_\tau Q)^2}$
are approximately constant in time,
and $\partial_\tau \lambda \approx -v^2$.
Both numerical simulations and qualitative analysis
indicate that there are two types of transition.  The
signature of a transition is that $v$ is not constant in time and
$\partial_\tau \lambda$ is not approximated by $-v^2$.  For
each type of transition a rule for $v$ is obtained by
considering an exact ``transition solution'' which approximates
the evolution.  One type of transition
is driven by the spatial curvature.  It has been studied in
various classes of spacetimes, including the Gowdy spacetimes,
and gives two consecutive Kasner spacetimes in
the BKL sequence.  In the Gowdy spacetimes
the oscillations eventually cease.  If $\partial_\tau \mu
\neq 0$ then a second type of transition occurs, in which
$\partial_\tau \mu$ grows in magnitude and decays again,
and throughout which the evolution remains Kasner,
but the Kasner directions rotate.
This rotation is geometrical.  A given Kasner direction may
be tangent to the symmetry orbit before the transition, but
not tangent after.  In the spatially homogeneous setting this
type of transition was noticed using Hamiltonian methods\cite{J}
(``centrifugal potential'') and has since been noticed
using dynamical systems methods.\cite{dynsys}

\section*{Acknowledgments}
This work was supported in part by US
National Science Foundation Grants PHY9800103 and PHY9800732.


\begin{thebibliography}{99}

\bibitem{bkl}V. A. Belinskii, I. M. Khalatnikov and E. M.
Lifshitz, \Journal{\em Adv.\ Phys.}{31}{639}{1982}.

\bibitem{pap1}M. Weaver, J. Isenberg and B. K. Berger,
\Journal{\PRL}{80}{2984}{1998}.

\bibitem{pap2}B. K. Berger, D. Garfinkle, J. Isenberg,
V. Moncrief and M. Weaver,
\Journal{\em Mod.\ Phys.\ Lett.}{A13}{1565}{1998}.

\bibitem{U(1)}B. K. Berger and V. Moncrief,  
\Journal{\PRD}{62}{123501}{2000}.

\bibitem{globalfol}B. K. Berger, P. T. Chru\'{s}ciel,
J. Isenberg and V. Moncrief,
\Journal{\em Ann.\ Phys.}{260}{117}{1997}.

\bibitem{pol}J. Isenberg and S. Kichenassamy,
\Journal{\em Jour.\ Math.\ Phys.}{40}{340}{1999}.

\bibitem{J}R. T. Jantzen,
in {\em Cosmology of the Early Universe}, ed.\
L. Z. Fang and R. Ruffini (World Scientific, Singapore, 1984).

\bibitem{dynsys}C. G. Hewitt, R. Bridson and  J. Wainwright,
gr-qc/0008037.

\end{thebibliography}
\end{document}